# Introducing Systems Thinking as a Framework for Teaching and Assessing Threat Modeling Competency

Siddhant S. Joshi[1], Preeti Mukherjee[2], Kirsten A. Davis[1], and James C. Davis[2]

School of Engineering Education, Purdue University[1]
School of Electrical and Computer Engineering, Purdue University[2]

**Abstract**
Computing systems face diverse and substantial cybersecurity threats. To mitigate these cybersecurity threats while developing software, engineers need to be competent in the skill of *threat modeling*. In industry and academia, there are many frameworks for teaching threat modeling, but our analysis of these frameworks suggests that (1) these approaches tend to be focused on component-level analysis rather than educating students to reason holistically about a system's cybersecurity, and (2) there is no rubric for assessing a student's threat modeling competency. To address these concerns, we propose using systems thinking in conjunction with popular and industry-standard threat modeling frameworks like STRIDE for teaching and assessing threat modeling competency. Prior studies suggest a holistic approach like systems thinking can be suitable for understanding and mitigating cybersecurity threats. Therefore, the purpose of this study is to develop and pilot two novel rubrics – one for assessing STRIDE threat modeling performance and the other for assessing systems thinking performance while conducting STRIDE.

To conduct this study, we piloted the two rubrics mentioned above to assess threat model artifacts of students enrolled in an upper-level software engineering course at Purdue University in Fall 2021, Spring 2023, and Fall 2023. Our results reveal that students who had both systems thinking and STRIDE instruction identified and attempted to mitigate component-level as well as systems-level threats. On the other hand, students with only STRIDE instruction tended to focus on identifying and mitigating component-level threats and discounted system-level threats. Our work contributes to the engineering education community by: (1) describing a new rubric for assessing threat modeling based on systems thinking; (2) identifying trends and blindspots in students' threat modeling approach; and (3) envisioning the benefits of integrating systems thinking in threat modeling teaching and assessment.

**Introduction**
With rapid developments in computer science and growing dependence on information technology, cybersecurity threats are evolving at a rapid rate [1], [2]. Cybersecurity is defined as the combination of technologies, resources, structure, and culture that is utilized to protect data in cyberspace and cyberspace-enabled systems from vulnerabilities, threats, exposure, and damages to ensure stability and sustenance [2, p. 2]. Further, as these cyber threats become more sophisticated, the industry needs to protect its systems against cybercriminals capable of penetrating their security systems [3]. For instance, the Siemens report [4] suggests that digitization has led to multiple cybersecurity challenges which if not addressed can lead to huge financial losses for the industry and society. Given the importance and potential damage that can be caused by cybersecurity threats, the responsibility to address these challenges relies on competent cybersecurity engineers.

The current state of the cybersecurity workforce suggests that engineers have a shortage of cybersecurity skills to address the security challenges that lie ahead [5]. For example, a survey by CSIS and McAfee in 2016 of IT decision-makers from eight countries indicates that 82% of employers felt that their workforce did not have the necessary cybersecurity skills and 71% believed that the skill gap caused measurable and direct damage to the security of their organizations [6]. Additionally, although the number of cybersecurity jobs has been increasing, almost 28% of these jobs nationally are still vacant as of 2023 [7]; mainly due to a lack of training on cybersecurity fundamentals and hands-on experience [3]. As a result, there is a demand for universities to educate students with fundamental competencies that help prepare them for addressing cybersecurity challenges. One such fundamental cybersecurity skill is threat modeling [8].

This paper presents the first step in our work and introduces a novel approach to teaching and assessing threat modeling based on principles of systems thinking. We begin this paper with the background and related work section where we identify the current gap in threat modeling teaching and assessment and highlight how system thinking can help with system-level threat modeling. Next, in the context section, we provide details of the software engineering course in which we situated our study and describe the changes we made to its current iteration by including a module on systems thinking and updating the course's threat modeling deliverable. After the context section, we discuss the methods of our study and introduce the two rubrics (one on systems thinking and the other on STRIDE) we developed for assessing threat modeling deliverables. Finally, we present the results of piloting our rubric on the course's threat modeling deliverable and discuss how systems thinking can be useful for threat modeling teaching and assessment.

**Background and Related Work**
Software engineering and secure software development have gathered attention because many of the cybersecurity threats arise due to defects in software [9]. Many industries still rely on fixing security flaws in software when a security situation arises [10]. Fixing security flaws in the software after a cyberattack often can be detrimental as losses might have already taken place. Additionally, as the attacks by malicious attackers continue and increase rapidly, knowledge and competencies associated with cybersecurity are essential during software development [11]. As a result, a security mindset should be developed when one learns the software development lifecycle [12]. Threat modeling is a crucial cybersecurity and secure software development skill [13] that helps analyze the risks associated with the software architectures and identify strengths and risks early on [14], [15]. Therefore, teaching threat modeling is important in a software engineering course to develop secure software.

*What is threat modeling?*
Threat modeling is a security analysis approach that involves assessing the applicability and relevance of threat scenarios that a system can face once it is deployed in the real-world environment [16]. It is a systematic approach to identifying, mapping, and mitigating design-level security problems (Soares Cruzes et al., 2018). It helps identify and describe security flaws, access points, and appropriate security requirements during the software development process to ensure that the software can be made capable of mitigating possible threats [17]. Threat modeling usually takes place during the early stages of the software development lifecycle [18], [19] as it helps fix issues during the development rather than rework the design after it has been deployed. Fundamentally, threat modeling involves identifying and understanding the architectural model of the system, identifying threats associated with each component of the system, and developing mitigation strategies to address the component-level threats [19]. Essentially, the process of threat modeling provides a structured way to develop secure software by allowing the developers to estimate the capabilities of the attackers based on the known threats faced by the system [20]. As threat modeling is a crucial skill for developing secure software, it is frequently taught to engineers during their undergraduate software engineering education.

*Frameworks Used for Threat Modeling*
There are various frameworks in place to practice and teach threat modeling. One of the most popular frameworks used for threat modeling is STRIDE. STRIDE is a threat modeling methodology where the engineer classifies the vulnerabilities of the system into six categories - Spoofing, Tampering, Repudiation, Information Disclosure, Denial of Service, and Elevation of Privilege [21]. Additionally, this approach also involves creating an architecture overview using data flow diagrams (DFD), decomposing the architecture into components, identifying threats affecting each component, and documenting and ranking the threats [13]. Approaches like DREAD (Damage, Reproducibility, Exploitability, Affected Users, Discoverability) are used in tandem with the STRIDE framework to rate, compare, and prioritize the severity of risk presented by each threat. In the end, the STRIDE framework results in the formulation of mitigation strategies targeted to address the identified threats.

Process for Attack Simulation & Threat Analysis (PASTA) threat modeling methodology [22] is also another popular framework for threat modeling. The advantage of the PASTA framework is that it combines business and technical objectives together [23] whereas STRIDE is focused only on the technical objectives. PASTA is a risk-centric framework where the engineer conducts threat modeling at a strategic level by involving key decision-makers in the organization. The PASTA framework requires the engineer to think from the perspective of the attacker similar to the STRIDE framework. The end result of the PASTA framework is that it produces an asset-oriented output where the organization can derive the impact of threats using simulated attacks [24].

Along with STRIDE and PASTA, other popular methods for threat modeling include OCTAVE (Operationally Critical Threat, Asset, and Vulnerability Evaluation) and Attack Tree techniques [22]. The OCTAVE method is used to assess mission-level threats from an organizational standpoint and does not focus on technological risks. The Attack Tree method is more suitable for architecture risk analysis, especially in complex situations, and may be an overkill for simple, familiar, or fully understood attacks [25]. The Attack Tree method is also used in combination with popular techniques such as STRIDE [26] and PASTA [23], [24]. LINDDUN, another threat modeling method uses a threat tree catalog that encompasses a wide variety of known and common attack paths or access points for each threat category such as Linkability, Identifiability, Non-repudiation, Detectability, Information Disclosure, Content Unawareness, and Non-compliance. It is often used early in the software development phase and employed to mitigate privacy threats associated with the software [27].

Although multiple threat modeling frameworks are available, STRIDE is often preferred over other methods in software development because it is considered the most mature threat modeling framework [23]. Further, STRIDE is frequently used in the industry [28] and is preferred in the leading secure software development processes [14]. In summary, STRIDE is the most popular and mature framework available to practice and thereby teach threat modeling to software engineering students.

*Drawbacks of Current Threat Modeling Approaches*
Though the above threat modeling frameworks are useful for engineers to build secure software, they have their set of drawbacks. In many of the above threat modeling frameworks, there is a lack of emphasis on relationship-based threat modeling [29]. This is because the majority of threat modeling techniques consider only the component-level threats faced by the system which hampers identifying scenarios related to the emergent threats which may arise due to interaction between different components of a system. [29], [30]. For instance, the STRIDE, the most popular threat modeling framework, does not account for the interaction between components as it aims to individually immune components susceptible to known threats. As a result, it fails to account for threats that may materialize when components of a system are connected with each other [21]. Furthermore, prior research in systems engineering shows that decomposing a system into components and analyzing each component separately (as done in STRIDE and other threat modeling frameworks) limits the solution designers' ability to understand how the overall system behaves [31], [32]. Hence, along with component-level analysis, threat modeling frameworks need to incorporate system-level threat analysis as well. Currently, to the best of our knowledge, none of the approaches used in threat modeling address threats that arise due to relationships between components or system-level threats.

Another drawback of the current threat modeling approaches is that they can lead to threat explosion upon software deployment. Threat explosion is a result of the growing complexity of threats that arise when a new component is added and the number of threats drastically increases

[33]. As a result, new system-level threats may arise as new components or services are added to the software. This further necessitates a need for a framework for system-level threat modeling. Additionally, changes in the system during software development may require revaluating all threats because relationships between components may change [29]. This may lead to a lot of rework in the early stages of the software development cycle.

Due to the above reasons, there is an opportunity for educators to teach and assess threat modeling that emphasizes not only the component-level threats but also prioritizes system-level threats arising from relationships between components. Systems thinking is one such thinking approach that aims to understand the dynamic nature of a system, its interconnections, synergy due to interconnections, and observe and predict the behavior of the system as a whole instead of just focusing on its parts [34], [35]. Therefore, in the next subsection, we propose systems thinking as a framework for teaching and assessing the threat modeling competency of software engineering students.

*Using systems thinking to supplement existing threat modeling frameworks*
As discussed earlier, in cybersecurity threat modeling, the software engineer needs to ensure that not only the component-level threats are addressed but so are the system-level threats that arise due to interconnections between different components of the system. Systems thinking is a skill that focuses on understanding the systemic properties of a system while also accounting for emergent trends arising from the combination of the connected parts [36]. There are multiple definitions and frameworks available for systems thinking from a variety of disciplines like engineering [34], [35], management [37], cognitive sciences [38], etc. However, we chose the definition and framework developed by Cabrera and Cabrera [39] because their framework helps develop a mental model needed to practice systems thinking [36]. Further, their framework has previously been used in educational contexts and is universally applicable to individuals with varying disciplinary backgrounds [36]. Cabrera and Cabrera [39] defined systems thinking as a four-part cognitive skill consisting of tenets like making distinctions (D), organizing the system (S) into parts and wholes, recognizing relationships (R) between parts and wholes of the system, and taking multiple perspectives (P). Taken together, this four-part skill helps develop a holistic approach to designing a solution to a problem.

In the context of threat modeling, the systems thinking approach translates as a way to help engineers account for and mitigate system-level threats that arise as a result of interaction between the components without discounting threats posed to each component individually. For instance, by practicing systems thinking during threat modeling, the software engineer can not only address threats arising due to the relationship between components but also consider threats from the perspectives of inside and outside attackers. Therefore, teaching systems thinking in conjunction with popular threat modeling techniques like STRIDE can help overcome the drawbacks of existing threat modeling techniques.

Although systems thinking has the potential to ensure more robust threat modeling practice, no prior work has looked to develop a STRIDE rubric and use systems thinking in combination with STRIDE to teach and assess threat modeling of software engineering students. Further, previous research on threat modeling does not provide evidence if popular approaches such as STRIDE foster system-level threat modeling while teaching threat modeling. Given the rising cybersecurity risks and the utmost importance of effective system-level threat modeling techniques, the purpose of our work is to propose systems thinking as a supplementary framework to use alongside STRIDE for teaching and assessing the threat modeling competency of upper-level software engineering students. Based on the above-discussed gaps, we aim to address the following research question:

Research Question - To what extent do upper-level software engineering students with and without systems thinking instruction practice systems thinking while applying the STRIDE threat modeling framework?

**Context**
*Course Description and Cybersecurity Aspects*
We situated our study in a three-credit software engineering course offered to Electrical and Computer Engineering (ECE) majors at Purdue University, a large Midwestern university in the USA. This course is offered in a synchronous modality to $3^{rd}$- and $4^{th}$-year bachelor's students in the ECE department. Enrollment of the course is 75-150 students per offering. The prerequisite knowledge is a two-course sequence in introductory programming and a course in data structures and algorithms. The learning outcomes cover software engineering methodologies (e.g., iterative vs. plan-based) and specific techniques for software design, implementation, validation, deployment, and maintenance. Pertinent to this study, one learning outcome relates to cybersecurity analysis.

The course uses a project-based learning approach to teach these outcomes. Students work in teams (groups of 3 to 4 individuals) on a semester-long software engineering project. Teams must provide weekly updates, but these are intended to help course staff assist struggling teams rather than as assessment instruments. The primary assessable assignments are the major milestones of the project – deliveries in week 4, week ~8, and week 16. The project requirements have been similar in all offerings of the course (Fall 2021, Spring 2023, Fall 2023). Teams build a replica of the NPM package registry and deploy it to a cloud platform (Google Cloud or AWS). Students in the course are allowed to use Large Language Models (LLMs) for assignments and team projects, including the threat modeling component. A more detailed description of the course and LLM usage is available in [40] and [41] respectively.

The previous iterations of this course (Fall 2021 and Spring 2023) taught threat modeling using the STRIDE framework. The instructor observed that the students' analyses were somewhat naïve, lacking a holistic perspective. In the Fall 2023 iteration of the course, the instructor took three steps to improve the student's threat modeling skills. These steps were:

1. *Training:* Additional learning material on systems thinking was developed and integrated into the learning module on threat modeling. This material was taught during one lecture and introduced the DSRP principles of systems thinking. The material described DSRP as offering a mindset with which to think systematically about a computing system while conducting a security analysis.
2. *Tooling:* The instructor worked with a company called ThreatModeler (https://threatmodeler.com/) to obtain educational access to the ThreatModeler platform. Students were able to develop their system models on this platform. Given a model, this platform performs automated threat identification (following a checklist) and recommends mitigations. The instructor emphasized that this automated support was simplistic and that if students relied on it then their system would be insecure.
3. *Assessment:* The instructor updated the assignment associated with the security analysis to ask students to provide more detailed rationales for their system models, identified threats, and mitigations. This enabled us to better assess the level of holistic thinking demonstrated by the students. These changes are detailed next.

**Framework constructs for rubric development**
*STRIDE*
STRIDE, a robust threat modeling framework developed by Microsoft [20], [21], encompasses six key types of security threats: Spoofing Identity (S), Tampering with Data (T), Repudiation (R), Information Disclosure (I), Denial of Service (DoS) (D), and Elevation of Privilege (E). Spoofing involves deceptive practices to assume the identity of trusted entities. Tampering threats revolve around unauthorized data modifications and compromising integrity. Repudiation pertains to individuals denying involvement in actions, and challenging accountability. Information Disclosure involves unauthorized access to confidential data. Denial of Service disrupts system availability through resource overload. Elevation of Privilege seeks unauthorized escalation of user privileges. This framework provides a comprehensive approach to identifying and addressing security threats, aiding security professionals in evaluating and mitigating potential risks. The components of the STRIDE framework discussed above act as the important constructs that our rubric (discussed in methods) needs to assess the STRIDE threat modeling performance.

*DSRP Systems Thinking.*
In the educational and research context, the Distinction – System – Relationship – Perspective (DSRP) is a systems thinking framework developed by Cabrera and Cabrera [39] that has been used to teach and learn systems thinking. The DSRP criteria have been intricately woven into the assessment process, offering a unique lens for evaluating an individual's systems thinking approach. The Distinctions (D) component ensures a keen eye for detail, recognizing nuances and differences between components and systems. Systems (S) thinking allows for a holistic evaluation, by understanding the role of each component of the system and how they contribute to the overall coherence of the system as a whole. Relationships (R) are scrutinized to assess how various components interact and how their interaction influences the system. Perspectives (P) demand a nuanced understanding of different points of view to observe and understand the system under study. Taken together, these four cognitive rules help evaluate how one demonstrates systems thinking. The tenets of the DSRP framework shown above act as the important constructs that our system thinking rubric needs (rubric introduced in methods) to assess in the context of threat modeling. We discuss how these constructs are used to develop the rubric in the methods subsection on rubrics.

**Methods**
This paper is part of an ongoing project to investigate how systems thinking can be used in combination with popular threat modeling frameworks like STRIDE to teach and assess component-level and system-level threat modeling to upper-level software engineering students. In this section, we provide an overview of the methods we used in our study. We begin by describing the software engineering course where we piloted our study. Next, we discuss our data collection strategy, introduce the pilot version of our rubric, our data analysis approach (scoring strategy using our rubric), and ethical considerations.

*Data collection*
To answer our research question, we collected data on the students' team projects. In the project, student teams had to deliver the implementation of the software they developed and communicate the final status of the project.

Specifically, the project's final milestone (week 16) asks student teams to describe and provide evidence of the achieved functional requirements (e.g. the system is deployed to Amazon Web Services and supports the requisite API) and non-functional requirements, notably a security case that includes a security analysis using STRIDE. For this study, we analyzed the security case provided by each team.

In the security case deliverable of the project, the student teams had a baseline requirement to discuss the following various components of their threat model (1) systems data flow diagram with trust boundaries (2) threat model(s) (3) Consideration of STRIDE-type threats in the context of system and threats, possibly with reference to OWASP Top 10 and other lists of security best practices and threats, (4) Mitigations taken in response to the analysis, and (5) risks they did not mitigate along with their rationale. Figure 1 provides an overview of the security analysis prompt of the team project.

The same deliverable was common to all student teams - Fall 2021, Spring 2023, and Fall 2023. In Fall 2023 however, we tweaked the response template to include new prompts focused on systems thinking. Figure 2 shows the new prompts we added in the Fall 2023 to the response template. Our objective behind adding the prompts on systems thinking was to understand if student teams practiced DSRP principles of systems thinking while conducting security analysis using STRIDE. Specifically, we wanted to understand if teams thought about, identified, and considered aspects like the relationship between components, threats arising from the relationship between components, threats from multiple attacker perspectives, and system-level threats while performing security analysis using STRIDE.

**Figure 1**
*The figure describes a security case as a deliverable for student team projects. The security case focuses on developing a threat model based on STRIDE principles for the ACME Corporation and suggesting mitigation for the threats identified. Further, the figure mentions the baseline requirements that each team had to prepare for their security case.*

*Security requirements*
ACME Corporation would like to restrict access. Your system should support:

- Register, authenticate, and remove users with distinct "upload", "search", and "download" permissions.
- The system should permit authentication using a combination of username + secure password and yield a token to be supplied to the other APIs as a payload parameter.
- This token should remain valid for 1000 API interactions or 10 hours.
- Some users should have an "admin" permission. Only administrators can register users. Users can delete their own accounts.
- Some teams at ACME Corporation have experimental code that should not be accessible by other members of the company. There should be a "user group" mechanism that can be specified at user creation time. During upload, a package can be marked as "secret". Such a package can only be queried by a member of the group that uploaded it.

You must persuade ACME Corporation that your system does not expose them to substantial security risks. To this end, you should prepare a security case in three parts:
1. *Design for security:* Develop a system design in the ThreatModeler platform.
2. *Analyze and mitigate risks*:
    a. Check-box: Handle all the "best practices" recommendations provided by ThreatModeler.
    b. Principled: Conduct a security analysis based on STRIDE in your system.
3. *Demonstrate:* Carry out automated security probing of your system using the RESTler tool. Document and repair any detected vulnerabilities. Indicate how up to two of them came to enter your system (perhaps with a "Five Whys" analysis). (To do this, you will need reasonable record-keeping in your version control system – traceability from code to author and associated processes and decisions)

ACME Corporation would like to be able to conduct security audits of packages in the registry. They should be able to request the audit of one package, a group of packages, or the entire registry. As a default, they would like to use the "audit" mechanism provided by npm. They would also like to be able to run audits using an arbitrary JavaScript program. This JavaScript program expects to run under Node.js v18, and accepts one command line argument: "ZIP_FILE_PATH AUDIT_FILE_DIR". Audit results will be written to a file in AUDIT_FILE_DIR. The result for all audited packages should be displayed. For any package on which the program exits with a non-zero code, the results of the audit program should be available.

*Baseline requirement*
All teams must prepare the security case. Here are the key parts of the case.
- Dataflow diagram with trust boundaries
- Threat model(s)
- Consideration of STRIDE-type threats in the context of your system and threats, possibly with reference to OWASP Top 10 and other lists of security best practices and threats
- Mitigations you took in response to the analysis.

**Figure 2**
*New prompts on systems thinking that we added to security case deliverable in Fall 2023. The previous iterations of the course in Fall 2021 and Spring 2023 did not explicitly ask students to discuss risks emerging from interactions.*

> **Risks resulting from component interactions**
>
> The STRIDE framework (per Microsoft) advises you to divide and conquer – analyze each component in turn. You have now done that.
>
> Are there any instances in your system where a risk emerges from the interaction of multiple components? If you can, identify one case and describe it. Did you find it during your STRIDE process or only just now?
>
> Are there instances in your system where the requirements of one component (e.g., security, correctness, performance) may negatively affect the security requirements of another component or the system? If you can, identify one case and describe it.
>
> **Root cause analysis**
>
> Presumably (1) you did not intentionally create any security vulnerabilities, yet (2) found some through this process. Choose two interesting vulnerabilities. Describe why they happened.

*Rubric*
In this subsection, we introduce a rubric we developed based on STRIDE and system thinking frameworks for threat modeling.

*Need for a rubric*
Rubrics provide a guide for scoring student work and help assess the performance of a particular learning outcome [42]. Rubrics act as a scoring tool that helps evaluate the student on specific dimensions of the assignment by providing a detailed explanation of what constitutes a satisfactory and unsatisfactory level of performance on the assignment [43]. Prior research suggests that well-developed rubrics can aid in evaluating student performance in a reliable way [44]. As a result, rubrics are a useful, effective, and reliable tool to understand and assess if the student's work has met a sufficient level of satisfactory performance on the learning outcomes of a given assignment.

As discussed previously, no prior work has developed a rubric for STRIDE as well as systems thinking to assess the threat modeling performance of software engineering students. Therefore, we have developed an initial and pilot version of two analytical rubrics – one to assess STRIDE performance, and the second to assess systems thinking performance. We developed an analytical rubric because it helps assign a numeric score to a specific construct or learning outcome being measured based on the quality of the response provided in the assignment [45]. In our case, these learning outcomes are called constructs. These constructs include various steps in the STRIDE and DSRP systems thinking frameworks. We introduced these constructs in the frameworks section and now discuss them below.

*Constructs of the rubric*

Based on the STRIDE and DSRP constructs discussed in the frameworks section, we have developed two rubrics that help us understand how well students perform on the different learning outcomes of STRIDE and DSRP. We broke down the rubric into specific assessment constructs for (1) STRIDE and (2) DSRP. Each of the constructs of STRIDE and DSRP was divided into three scales based on the quality of student response: Beginner, Intermediate, and Advanced. This thorough approach can help us see how good students are at handling security issues using STRIDE as well as how they demonstrate systems thinking skills like making distinctions, understanding systems, recognizing relationships, and viewing the threat model from multiple perspectives. By looking closely at these constructs and scales, we introduce a detailed assessment rubric that will help educators evaluate STRIDE and systems thinking performance. The STRIDE rubric and DSRP rubric for threat modeling are shown in Appendix A and their sample versions are shown in Table 1 and Table 2 respectively. Further, we divided the STRIDE rubric into two phases (1) Modeling and (2) Threat Analysis. The modeling phase consisted of constructs of *Defining threats in a model, Defining security requirements, Dataflow diagram (DFD), and Documentation.* The threat analysis phase consisted of *Spoofing, Tampering, Repudiation, Information Disclosure, Denial of Service, Elevation of Privileg*e, and *Mitigation Strategy.*

**Table 1**

*A sample version of the STRIDE rubric to assess the security case deliverable is shown. The detailed version of this rubric is available in the Appendix. Note: A score of 0 was given if there was no response related to a given construct*

| Constructs | Beginner (score = 1) | Intermediate (Additional to beginner skillset; Score = 2) | Advanced (Additional to intermediate skillset; Score = 3) |
|---|---|---|---|
| *Defining threats in a model* | Identifies if a system has a potential source of threat or not. | Understands the threat that contributes to the risk, and the extent of how the threat impacts the components. | Identify the specific threat agents that can harm the components and/or the system. |
| *STRIDE - Spoofing* | Defines Spoofing and identifies which part of the model contributes to the same | Identifies the property(ies) violated: E.g., Authentication | Identifies the extent to which the threat is affecting the component and/or system and ranks its importance. |
| *Mitigation strategies* | Identifies correct mitigation strategies based on the properties violated | Discusses the effectiveness of mitigation strategies and specifically answers the questions: Which threat? What strategy? How to implement it? And, why this implementation mitigates the earlier mentioned problem? | 1. Identifies the threats that exist due to the interaction between the different components and their extent of contribution to the threat of the entire model. Discuss possible mitigation strategies.<br>2. Discusses about efficient implementation of mitigation strategies and talk about resource-constrained situations.<br>3. Discusses possible scope of trade-offs and if there are new requirements in the system for threat mitigation. |

**Table 2**

*A sample rubric was developed to assess DSRP systems thinking for the security case deliverable. The detailed version of this rubric is available in the Appendix. Note: A score of 0 was given if there was no response related to a given construct*

| DSRP Construct | Beginner (score = 1) | Intermediate (Additional to beginner skillset; Score = 2) | Advanced (Additional to intermediate skillset; Score = 3) |
|---|---|---|---|
| *Distinctions* | Identifies basic distinctions between components and system. | Analyzes distinctions in moderately complex threat scenarios. | Critically evaluates and synthesizes complex distinctions between different threats for system and components, showcasing depth. |
| *Relationships* | Identifies basic relationships between components. | Analyzes relationships between components in a nuanced manner. | Evaluates intricate relationships between components as well as how they collectively contribute to the system, demonstrating advanced insights. |

*Participants*

To answer our research questions, we collected data from 24 student teams in Fall 2021, 37 student teams in Spring 2023, and 18 student teams in Fall 2023 who were enrolled in the software engineering course at Purdue University, USA. The students enrolled in this course are junior and senior year students from either electrical or computer engineering majors. As discussed, the Fall 2021 students and Spring 2021 students did not receive any instruction on systems thinking whereas students from Fall 2023 did.

For this study, we randomly selected five student team projects from each of the Fall 2021, Spring 2023, and Fall 2023 semesters (total 15 projects) and analyzed their security case deliverables using the rubrics we developed. Next, we discuss our scoring approach based on the rubric we developed for assessing STRIDE and system thinking during threat modeling.

*Data analysis*
*Scoring using the rubric.*
The first step in the data analysis process was to score the security case deliverable using the rubrics shown above. For each dimension, students' responses were scored on a scale of 0–3 We used a score of zero for no response, one for beginner-level response, two for intermediate-level response, and three for advanced-level response if teams successfully fulfilled the scoring requirements of the given constructs of STRIDE and DSRP systems thinking as shown in the above rubrics.

The scoring took place in two stages. In the first stage, the two members of the research team met and selected five student projects at random and scored their security case based on the initial version of the rubric. After the first stage of scoring, the two members identified irregularities and came to a consensus on consistent definitions and interpretation for each construct as well as each score assigned in the two rubrics. To ensure consistency in the interpretation and scoring of the rubric, they added clarificatory sentences to each score of the rubric. In this way, a refined rubric was developed. This rubric was shared with the course instructor and an engineering education faculty for feedback, both of whom are authors of this paper.

In the second stage, two members of the research team scored the remaining security case deliverables from Fall 2021, Spring 2023, and Fall 2023 using the refined rubric. Once all the scoring was completed, the two members met and came to a consensus on the final score given to each construct of the rubric. The average rating agreement was close to 94%.

After the analysis was completed, we compared how the students from Fall 2021, Spring 2023, and Fall 2023 performed on their systems thinking (DSRP) and STRIDE framework scores (modeling and threat analysis phases). The comparison of scores for each STRIDE and systems thinking construct using histograms and averages helped us address our research question.

*Ethical considerations*
The Purdue University IRB has reviewed our study as an analysis of existing data because the team project reports were collected as part of the assessment for the software engineering course (Purdue IRB #2024-120). To ensure anonymity, we masked student names and deidentified the data by assigning a participant ID to their project submission. The participant key and the deidentified data were stored on a secured cloud drive and the original identifiable data were deleted upon deidentification. Additionally, to avoid coercion, the team projects were analyzed using the rubric only after the end of the Fall 2023 semester once all grades had been submitted.

**Results**

In the previous section, we introduced our proposed rubric for assessing STRIDE and systems thinking performance during threat modeling, and, in this section, we will describe the results of piloting this rubric. The purpose of this section is to compare STRIDE and systems thinking performance of three groups during threat modeling – one who received instruction on systems thinking and STRIDE (Fall 2023) and the other two who received only instruction on STRIDE (Fall 2021 and Spring 2023). By making the comparison we aim to answer the following research question: To what extent do upper-level software engineering students with and without systems thinking instruction practice systems thinking while applying the STRIDE threat modeling framework? The comparison will help us, and other educators (1) draw preliminary inferences on the using systems thinking to supplement threat modeling and (2) determine if it is promising to explore the intersection between threat modeling and system thinking in the future.

First, we share the results of scoring using the rubric and discuss how students from Fall 2023 compare with Spring 2023 and Fall 2021 on their STRIDE analysis and systems thinking performance. Second, we explain qualitative differences between student teams' performance on DSRP systems thinking by comparing responses of two student teams' security case deliverable. Third, we highlight the trends and blind spots related to systems thinking and STRIDE analysis that we noticed while scoring using the rubric. The preliminary results of our investigation are discussed below.

### *Results of scoring using STRIDE and systems thinking rubric*

Our scoring results (shown in Figure 3 and Figure 4), reveal that most of the student teams from Fall 2021 (group 1), Spring 2023 (group 2), and Fall 2023 (group 3) have done well in developing their threat models using the STRIDE framework. We observe that the average scores of all three groups in the modeling and threat analysis phases of STRIDE (Figure 3) are close by and consistent (ranging from 2.6 to 3). Further, most of the teams from the three groups have scored a 2 or 3 on their STRIDE modeling and threat analysis phases. In some cases, student teams have scored zero on STRIDE properties because they did not discuss components and affected security properties associated with components that might be susceptible to the threats.

For scores associated with the systems thinking performance of students (Figure 4), we observe that Fall 2021 and Spring 2023 students demonstrated beginner to intermediate level (with average scores generally between 1.20 and 1.60 and most scores being either 1 or 2) systems thinking while developing their threat models. Observing the Fall 2023 students, we realize that they did very well on their systems thinking scores and generally demonstrated an intermediate to advanced level of systems thinking (average scores between 2.4 and 2.8 and most scores being either 2 or 3) while developing their threat models using STRIDE.

**Figure 3**
*This figure presents the scoring results as per the STRIDE rubric. The first figure here shows the average score received by the three groups in the Modeling and Threat Analysis phases. The second and third figure shows the number of times each group received a score of 0, 1, 2, and 3 during the Modeling and Threat Analysis phases.*

|  | Fall 2021 | Spring 2023 | Fall 2023 |
| --- | --- | --- | --- |
| Average Modeling score | 2.85 | 2.6 | 3 |
| Average Threat analysis score | 2.49 | 2.40 | 2.74 |

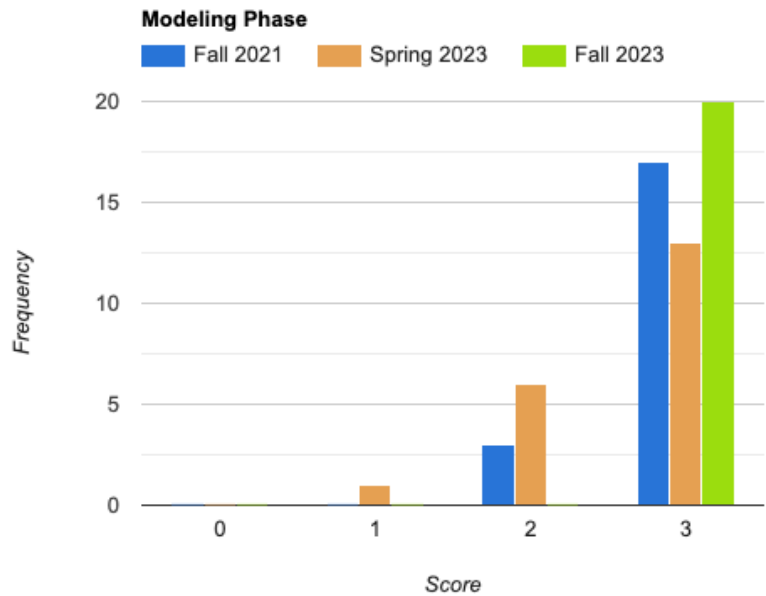

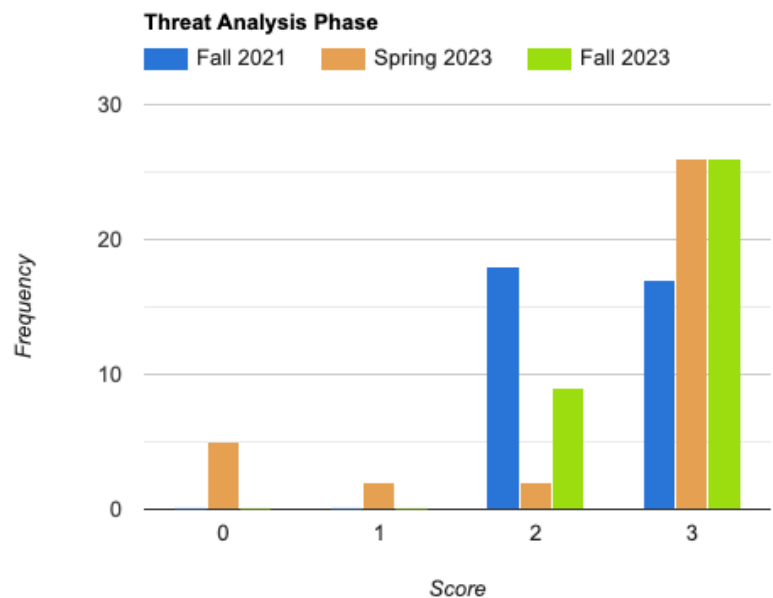

**Figure 4**
*This figure presents the scoring results as per the Systems thinking rubric. The first figure here shows the average score received by the three groups on each of the DSRP constructs. The second figure shows the number of times each group received a score of 0, 1, 2, and 3 during DSRP systems thinking.*

| Average | Fall 2021 | Spring 2023 | Fall 2023 |
|---|---|---|---|
| D | 1.60 | 1.40 | 2.60 |
| S | 1.20 | 1.40 | 2.40 |
| R | 1.40 | 1.40 | 2.80 |
| P | 2.40 | 1.60 | 2.40 |
| DSRP | 1.65 | 1.45 | 2.55 |

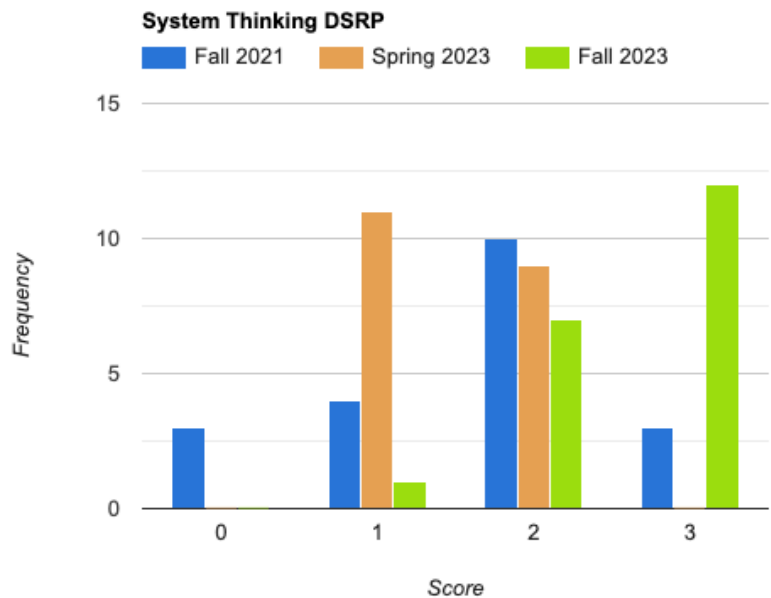

Taking the STRIDE and systems thinking results together, our results indicate that although Fall 2021 and Spring 2023 students performed well on STRIDE, they had beginner to intermediate-level system thinking performance. Further, the scoring of their threat modeling approach reflects their emphasis on component-level aspects of threat modeling rather than a systems-level perspective. On the other hand, as the Fall 2023 students had a module on systems thinking, they demonstrated relatively more advanced level of systems thinking and STRIDE performance while (higher average scores than the other two groups) performing the threat modeling. Additionally, the scoring results of Fall 2023 students indicate that their threat modeling approach shifted their focus from just component-level threat modeling toward more system-level threat modeling.

*Qualitative differences between students' systems thinking scores.*
As teams from all the semesters scored well on their STRIDE analysis, in this subsection, we qualitatively compare the systems thinking performance of Fall 2023 teams and teams from the other two semesters. To make this comparison, we have chosen examples from students' security case deliverable that focus on the Systems and Perspective constructs of the DSRP framework. For this comparison, we selected one team from the Fall 2021 semester (hereby referred to as 2021-Team A) and one from the Fall 2023 semester (hereby referred to as 2023 - Team U). These two teams were selected for comparison because they both did well on the STRIDE analysis but had very different scores on the DSRP system thinking constructs. The 2023 - Team U team had a score of three on each of the DSRP constructs whereas 2021-Team A had a score of one, zero, zero, and two on the D, S, R, and P constructs respectively.

Comparing the teams on their Systems construct, we observe that while developing a mitigation strategy for Spoofing, 2021-Team A recognizes and mitigates only basic component-level threats. For instance, to address the threat of intruders accessing repository files, they have implemented the following mitigation:

> "**Risk:** *Intruder gets access to our repository's files.*
>
> ***Mitigations applied****: only authenticated users like team members and people who were given access to the repo can view the repo content.*
>
> ***Degree of risk resolution****: High*
>
> ***Suggestions for additional mitigations, if needed****: N/A."*

On the other hand, 2023 - Team U while developing a mitigation strategy for Spoofing recognizes multiple threats, looks at multiple vulnerable components, and how their vulnerability can cause issues to the system. Their mitigation strategy for a risk associated with intruders gaining access is as follows:

> "**Risk***: Weak Identity, Credential, and Access Management*
>
> ***Mitigations applied****: Mitigations applied: Use of strong authentication mechanisms (MFA), centralized identity management (AWS IAM), adherence to least privilege principle, AWS CloudWatch monitoring. encryption of data in transit and at rest, and Role-Based Access Control (RBAC).*
>
> ***Degree of risk resolution****: All team members are required to enable multi-factor authentication with their AWS account. We use AWS IAM to assign roles to users to give them access to specific components within the system. We adhere to a least privilege principle along with RBAC to make sure that all components have access to the minimum needed levels. AWS DynamoDb and 53 automatically encrypt data at rest and utilize AWS KMS to make sure data is secure. Finally, through AWS Amplify we use HTTPS methods to transfer our data in an encrypted manner. With these mitigations in place, it greatly reduces the chances of unintended user access.*
>
> ***Suggestions for additional mitigations, if needed****: If we were not limited to AWS free tier, we could implement AWS X-Ray to view, filter, and gain insights into that data to identify issues and opportunities for optimization or could use AWS Shield advanced to protect our registry better"*

Observing the mitigation strategy of the two teams, we notice that 2023 - Team U demonstrates a deeper level of understanding of the threat related to accessing information than 2021-Team A. Subsequently, 2023 - Team U develops a mitigation strategy that has layers of security for each vulnerable component of the system while also placing mitigations measures at weak points of the system so that no unintended user gets access to the system. On the other hand, 2021-Team A does not describe their mitigation strategy in detail and only describes that they prevented access to a component but not the whole system. Further, 2021-Team A does not discuss 'how' they prevented access to the vulnerable component whereas 2023-Team U does by specifying mechanisms like MFA, least privilege principle, etc.

Now, comparing the teams on their Perspective construct, we notice that 2021-Team A identifies two trust boundaries and only mentions that there are threats from both insiders and outsiders. 2023 - Team U on the other hand identified not just the insider and outsider threats but also explained the rationale behind having multiple trust boundaries. In Figures 5 and 6m we see how the two teams describe the trust boundary and multiple perspectives through which they analyzed the threats to the system.

From the two figures, we notice that 2023 - Team U not only recognized how the attacks on the system can take place from multiple perspectives but also set up measures in the form of trust boundaries to counter those attacks. 2021-Team A on the other hand only mentioned that attacks can take place from multiple perspectives but did not discuss how the trust boundary helps prevent those attacks. Further, 2023 - Team U also discussed various attacker perspectives while developing their mitigation strategies during STRIDE while 2021-Team A did not discuss the role of different attackers while developing their mitigation strategies.

**Figure 5**
*This figure presents trust boundaries and attacker perspectives discussed by 2021-Team A in their security case deliverable.*

For each trust boundary indicated, describe the nature of the untrusted party involved (e.g. "outsider threat [e.g. external hacker]" or "insider threat [e.g. ACME employee with valid credentials]" or "infrastructure provider threat [GCP]").

**Trust boundary #**: 1
**Untrusted party**: Outsider threat

**Trust boundary #**: 2
**Untrusted party**: Insider threat

**Figure 6**

*This figure presents trust boundaries and attacker perspectives discussed by 2023 - Team U in their security case deliverable.*

For each trust boundary indicated in the DFDs, describe the nature of the untrusted party involved (examples: "outsider threat [e.g. external hacker]" or "insider threat [e.g. ACME employee with valid credentials]" or "infrastructure provider threat [AWS]").

---

**Trust boundary #: 1**
**Untrusted party**: User Input to Server
**Rationale for this boundary**: Preventing injection attacks and ensuring data integrity. Unvalidated user input can be a source of various vulnerabilities, including Cross-Site Scripting (XSS). Validating and sanitizing user input at this boundary helps ensure that the data processed by the server is safe and conforms to expected formats.

**Trust boundary #: 2**
**Untrusted party**: API Gateway to Backend Services
**Rationale for this boundary**: Protecting backend services from unauthorized access and malicious input. The API Gateway serves as a protective layer for backend services. Authenticating and authorizing requests at this boundary ensures that only valid and authorized requests reach the backend, minimizing the risk of abuse or unauthorized access.

---

**Trust boundary #: 3**
**Untrusted party**: IAM User Authentication
**Rationale for this boundary**: Protecting user accounts and sensitive information. This trust boundary is where user credentials are verified, and access privileges are granted. Ensuring secure authentication mechanisms and proper authorization controls at this boundary is vital for safeguarding user accounts and their associated data.

**Trust boundary #: 4**
**Untrusted party**: Build and Deployment Pipeline
**Rationale for this boundary**: Ensuring secure and reliable software delivery. The transition between development, staging, and production environments is a critical trust boundary. Securely managing deployment credentials, validating artifacts, and following secure deployment practices at this boundary help prevent unauthorized releases and ensure the integrity of the deployed software.

**Trust boundary #: 5**
**Untrusted party**: Logging and Monitoring
**Rationale for this boundary**: Protecting log data integrity and ensuring secure access. Logs and monitoring data contain sensitive information and play a crucial role in detecting and responding to security incidents. Protecting log data integrity, ensuring secure transmission, and restricting access to logs help maintain the confidentiality and reliability of monitoring information.

**Trust boundary #: 6**
**Untrusted party**: Serverless Functions
**Rationale for this boundary**: Managing secure execution of serverless code. Serverless functions operate within a defined environment, and the trust boundary is crucial for configuring secure function execution. This includes managing permissions, validating inputs, and ensuring the overall security of serverless deployments.

**Trust boundary #: 7**
**Untrusted party**: Third-Party Integrations
**Rationale for this boundary**: Managing external dependencies securely. Integrating with third-party services introduces a trust boundary to control access and interactions. Validating inputs, ensuring secure communication, and verifying the security practices of third-party services mitigate the risk of vulnerabilities arising from external dependencies.

***Trends and blind spots in students' threat modeling approach***

In this subsection, we discuss the trends and blind spots we noticed in teams' security case deliverables. We begin by discussing trends and blind spots we observed during STRIDE (Table 3) and systems thinking (Table 4).

*STRIDE*
**Table 3**
*This table described the trends and blind spots we observed in students' security case deliverable while scoring their projects using the STRIDE rubric.*

| Category | Observation |
|---|---|
| Trend | Most teams clearly defined the security requirements and threats associated with their software and hence, received an advanced score on these constructs |
| Blindspot | The majority of the Fall 2021 and Spring 2023 teams that received a score of two presented only a high-level overview of the dataflow diagram but did not present a detailed version of the dataflow diagram with all the trust boundaries and components. |
| Trend | Fall 2023 had a detailed DFD with all trust boundaries and components. They exhaustively discussed the problematic regions in the dataflow diagram and talked about potential vulnerability scenarios with respect to trust boundaries. We believe that by using the functionalities of the ThreatModeler tool students were able to build detailed DFD, and system models, and also identify risks associated with each component. |
| Blindspot | Fall 2021 and Spring 2023 teams did not exhaustively discuss the problematic regions in the dataflow diagram and talk about potential vulnerability scenarios for a component or a system based on its trust boundaries |
| Blindspot | A few teams in all three semesters did not discuss affected security properties during STRIDE risk analysis and/or mitigation and hence |
| Blind spot | Some teams from Fall 2021 and Spring 2023 received a score of one if they identified risks and mitigations but implemented only a few mitigations. Additionally, they received a score of two if they discussed component-level mitigation but did not discuss mitigations for situations where there were threats due to interactions between components |
| Trends | Fall 2023 students discussed different scenarios where components interacted and hence, may have identified the potential threats associated with interactions. Further, they also recommended mitigation strategies for a few threat scenarios that arose due to interaction between components |

*DSRP Systems thinking*
**Table 4**
*This table described the trends and blind spots we observed in students' security case deliverable while scoring their projects using the systems thinking rubric.*

| Category | Observation |
| --- | --- |
| Trend | Most teams from Fall 2023 teams identified system-level threats and/or emergent threats arising due to interaction between components and scored high on systems thinking constructs of Systems and Relationships |
| Trend | While discussing system-level threats, many of the Fall 2023 teams also discussed different scenarios where and how the system could be vulnerable |
| Blindspot | Teams from Fall 2021 and Spring 2023 either did not discuss system-level threats, or interaction threats, or discussed them very minimally (e.g., mentioned only in security requirements). Instead, they chose to focus only on component threats. These teams scored between beginner and intermediate on S and R |
| Blindspot | Fall 2021 and Spring 2023 teams did not exhaustively discuss the problematic regions in the dataflow diagram and talk about potential vulnerability scenarios for a component or a system based on its trust boundaries |
| Trend | Teams from Fall 2023 scored high on Distinctions because they drew detailed DFDs. We observed that the ThreatModeler tool functionalities helped students develop detailed DFDs and describe the role and risks associated with each component. They discussed specific and unique risks (both component and system level) for each STRIDE property and identified mitigation strategies for each risk. A few Fall 2021 and Spring 2023 teams did the same. |
| Blind spot | Although students from the Fall 2021 and Spring 2023 performed well with the DFD, they did not always describe the role of each component of the system and did not discuss distinct system-related threat or failure scenarios |
| Trends | Many teams across the three groups identified attackers from multiple perspectives i.e., insider attackers or outsider attackers. At times teams from Fall 2021 and Fall 2023 also described how at least one of the attackers would implement an attack |
| Trends | Many teams from Fall 2023 described not just an attacker and their implementation strategy but also presented how failure or threats can take place and the consequences that could happen at component and systemwide levels |

**Discussion**
The purpose of our paper was to use STRIDE and introduce systems thinking as frameworks for teaching and assessing the threat modeling competency of upper-level software engineering students. While teaching STRIDE threat modeling using a holistic systems approach, the instructor (1) incorporated modules on systems thinking; (2) acquired and provided access to the educational version of the ThreatModeler platform as a tool for developing system models; and (3) Updated the assessment on security analysis by asking students to provide more detailed rationale for their system models, identified threats, and mitigation strategies. To assess the security analysis in student team projects, we developed two separate rubrics, one for assessing their STRIDE threat modeling performance and the other for assessing systems thinking performance during STRIDE threat modeling. We piloted the above rubrics on two student groups – the first which had received instruction only on STRIDE threat modeling (Fall 2021 and Spring 2023), and the second which had received instruction on both STRIDE and systems thinking (Fall 2023). Our results revealed that Fall 2023 students who had received STRIDE and systems thinking instruction performed better than the Fall 2021 and Spring 2023 groups in holistically identifying and mitigating component as well as system-level risks. Given that students were holistically able to identify and work on mitigating system-level risks, we think that there is potential for using systems thinking in threat modeling teaching and assessment.

Currently, STRIDE is the industry standard and a popular framework for teaching threat modeling [23]. STRIDE is taught because it is easy to teach, not time-consuming, and can be incorporated into existing curriculum without many changes. Additionally, the resources for teaching STRIDE like its documentation and tools are readily available (e.g., see [Security Design using Microsoft STRIDE](#)). However, one of the drawbacks of STRIDE is that it is not originally developed to identify and mitigate system-level threats and threats arising due to the relationship between components [21]. Our results suggest that teaching a 1-week module on systems thinking in addition to STRIDE can help students identify system-level threats and also account for threats arising due to the relationship between components. This weeklong systems thinking plug-in module can be particularly useful because it is difficult to dedicate multiple weeks to teach other secure software development techniques like NIST Cybersecurity Framework [46] or NIST Secure Software Development Framework [47] in a software engineering (SWE) course.

From an assessment standpoint, our results suggest that incorporating systems thinking questions in student assignments of threat modeling can be useful in identifying trends and blind spots in the security case deliverables. Particularly, the blind spots we identified help reiterate that the STRIDE threat modeling framework does not probe SWE students to consider the relationships between components and their resultant emergent threats [21]. However, including probing questions on systems thinking during STRIDE can be advantageous as they prompt students to think about relationships between components and potential threat scenarios that arise due to the relationships. These modifications to the security case deliverable can help overcome the drawbacks of frameworks like STRIDE. Further, identifying relationships between components can be useful for identifying complex threat scenarios that arise when component interactions take place in a system [29]. Thus, there are potential benefits to incorporating systems thinking questions in student assignments on threat modeling.

Along with tweaking the assessment template, our work also focused on developing and piloting two rubrics, one for assessing STRIDE performance and another for systems thinking performance while conducting STRIDE. Our pilot results showed the potential for using these rubrics for assessing the STRIDE and systems thinking competencies during STRIDE. Our ongoing and future work will thus focus on building off these baseline results of the rubric and developing the rubric as per the procedure suggested by [48]. Specifically, we plan to longitudinally test student work for the remaining Fall 2021, Spring 2023, and Fall 2023 security case deliverables and identify any potential concerns associated with the rubric for further refinement. In addition, we will also make this rubric accessible to other professionals and educators and ask them to use the rubric to evaluate similar security case artifacts. We plan to seek their feedback and use their input to improve the construct validity of our rubric. Our end goal of this project is to develop teaching and assessment resources that combine threat modeling and systems thinking principles so that educators can use them to prepare SWE students for developing secure software.

**Conclusion**

This paper focuses on using STRIDE and introducing systems thinking as frameworks for teaching and assessing the threat modeling competency of upper-level software engineering students. As a part of this work, we introduced a module on systems thinking in a SWE course and developed two novel rubrics - one for STRIDE and another for systems thinking. We piloted these rubrics to analyze security case deliverable of SWE student team projects. Our results reveal that student teams who had systems thinking as well as STRIDE instruction were able to identify both system-level and component-level threats in their team projects and attempted to address them. Teams who received only STRIDE instruction tended to focus on component-level threats and minimally identified and mitigated system-level threats. Thus, we believe that there is potential for using systems thinking in threat modeling teaching and assessment. In summary, our work contributes to both teaching and assessment of the crucial skill of threat modeling to improve cybersecurity education and prepare students with the necessary skills for the workforce.

**Acknowledgements**
This work was funded by the Purdue Engineering Education Explorers Program, and by a pedagogy grant from the Elmore Family School of Electrical and Computer Engineering.

# References

[1] M. J. Assante, "Infrastructure Protection in the Ancient World," presented at the 2009 42nd Hawaii International Conference on System Sciences, IEEE Computer Society, Dec. 2009, pp. 1–10. doi: 10.1109/HICSS.2009.775.

[2] A. Hussain, A. Mohamed, and S. Razali, "A Review on Cybersecurity: Challenges & Emerging Threats," in *Proceedings of the 3rd International Conference on Networking, Information Systems & Security*, in NISS '20. New York, NY, USA: Association for Computing Machinery, May 2020, pp. 1–7. doi: 10.1145/3386723.3387847.

[3] J. A. Lewis and W. Crumpler, "The Cybersecurity Workforce Gap," Center for Strategic and International Studies, Jan. 2019. Accessed: Dec. 06, 2023. [Online]. Available: https://www.csis.org/analysis/cybersecurity-workforce-gap

[4] Siemens, "Cybersecurity in the Modern Industrial World," *Harvard Business Review*, Feb. 07, 2019. Accessed: Dec. 06, 2023. [Online]. Available: https://hbr.org/sponsored/2019/02/cybersecurity-in-the-modern-industrial-world

[5] D. Bendler and M. Felderer, "Competency Models for Information Security and Cybersecurity Professionals: Analysis of Existing Work and a New Model," *ACM Trans. Comput. Educ.*, vol. 23, no. 2, pp. 1–33, Jun. 2023, doi: 10.1145/3573205.

[6] K. Trimlin and J. A. Lewis, "Hacking the Skills Shortage: A Study of the International Shortage in Cybersecurity Skills," McAfee and CSIS, Santa Clara, CA, 2016. [Online]. Available: https://www.mcafee.com/enterprise/en-us/assets/reports/rp-hacking-skills-shortage.pdf.

[7] CyberSeek, "Cybersecurity supply/demand heat map," Cyberseek.org. Accessed: Dec. 06, 2023. [Online]. Available: https://www.cyberseek.org/heatmap.html

[8] W. Xiong and R. Lagerström, "Threat modeling – A systematic literature review," *Comput. Secur.*, vol. 84, pp. 53–69, Jul. 2019, doi: 10.1016/j.cose.2019.03.010.

[9] X. Yuan, L. Yang, B. Jones, H. Yu, and B.-T. Chu, "Secure Software Engineering Education: Knowledge Area, Curriculum and Resources," *J. Cybersecurity Educ. Res. Pract.*, vol. 2016, no. 1, Jun. 2016, [Online]. Available: https://digitalcommons.kennesaw.edu/jcerp/vol2016/iss1/3

[10] S. Acharya and W. W. Schilling, "Infusing software security in software engineering," in *2017 ASEE Annual Conference & Exposition*, Columbus, OH, 2017.

[11] I. A. Buckley, J. Zalewski, and P. J. Clarke, "Introducing a cybersecurity mindset into software engineering undergraduate courses," *Int. J. Adv. Comput. Sci. Appl.*, vol. 9, no. 6, 2018.

[12] H. Gonzalez, R. Llamas-Contreras, and C. Guerra-García, "Cybersecurity Practices At The Initial Stages Of The Software Engineering Process," in *2021 9th International Conference in Software Engineering Research and Innovation (CONISOFT)*, Oct. 2021, pp. 219–226. doi: 10.1109/CONISOFT52520.2021.00037.

[13] I. Williams and X. Yuan, "Evaluating the effectiveness of Microsoft threat modeling tool," in *Proceedings of the 2015 Information Security Curriculum Development Conference*, in InfoSec '15. New York, NY, USA: Association for Computing Machinery, Oct. 2015, pp. 1–6. doi: 10.1145/2885990.2885999.

[14] R. Scandariato, K. Wuyts, and W. Joosen, "A descriptive study of Microsoft's threat modeling technique," *Requir. Eng.*, vol. 20, no. 2, pp. 163–180, Jun. 2015, doi: 10.1007/s00766-013-0195-2.

[15] W. Xiong, E. Legrand, O. Åberg, and R. Lagerström, "Cyber security threat modeling based on the MITRE Enterprise ATT&CK Matrix," *Softw. Syst. Model.*, vol. 21, no. 1, pp. 157–177, Feb. 2022, doi: 10.1007/s10270-021-00898-7.

[16] D. Van Landuyt and W. Joosen, "A descriptive study of assumptions in STRIDE security threat modeling," *Softw. Syst. Model.*, vol. 21, no. 6, pp. 2311–2328, Dec. 2022, doi: 10.1007/s10270-021-00941-7.

[17] M. T. J. Ansari, D. Pandey, and M. Alenezi, "STORE: Security Threat Oriented Requirements Engineering Methodology," *J. King Saud Univ. - Comput. Inf. Sci.*, vol. 34, no. 2, pp. 191–203, Feb. 2022, doi: 10.1016/j.jksuci.2018.12.005.

[18] P. Bedi, V. Gandotra, A. Singhal, H. Narang, and S. Sharma, "Threat-oriented security framework in risk management using multiagent system," *Softw. Pract. Exp.*, vol. 43, no. 9, pp. 1013–1038, 2013, doi: 10.1002/spe.2133.

[19] D. Dhillon, "Developer-Driven Threat Modeling: Lessons Learned in the Trenches," *IEEE Secur. Priv.*, vol. 9, no. 4, pp. 41–47, Jul. 2011, doi: 10.1109/MSP.2011.47.

[20] F. Swiderski and W. Snyder, *Threat modeling*. Redmond, Wash: Microsoft Press, 2004.

[21] S. Hernan, S. Lambert, T. Ostwald, and A. Shostack, "Uncover Security Design Flaws Using The STRIDE Approach." Accessed: Oct. 27, 2023. [Online]. Available: https://learn.microsoft.com/en-us/archive/msdn-magazine/2006/november/uncover-security-design-flaws-using-the-stride-approach

[22] A. Konev, A. Shelupanov, M. Kataev, V. Ageeva, and A. Nabieva, "A Survey on Threat-Modeling Techniques: Protected Objects and Classification of Threats," *Symmetry*, vol. 14, no. 3, Art. no. 3, Mar. 2022, doi: 10.3390/sym14030549.

[23] N. Shevchenko, T. A. Chick, P. O'Riordan, T. P. Scanlon, and C. Woody, "Threat modeling: a summary of available methods," Carnegie Mellon University Software Engineering Institute Pittsburgh United …, 2018.

[24] T. UcedaVelez, "Real world threat modeling using the pasta methodology," *OWASP App Sec EU*, 2012.

[25] V. Saini, Q. Duan, and V. Paruchuri, "Threat modeling using attack trees," *J. Comput. Sci. Coll.*, vol. 23, no. 4, pp. 124–131, 2008.

[26] A. Marback, H. Do, K. He, S. Kondamarri, and D. Xu, "A threat model-based approach to security testing," *Softw. Pract. Exp.*, vol. 43, no. 2, pp. 241–258, 2013, doi: 10.1002/spe.2111.

[27] K. Wuyts, L. Sion, and W. Joosen, "LINDDUN GO: A Lightweight Approach to Privacy Threat Modeling," in *2020 IEEE European Symposium on Security and Privacy Workshops (EuroS&PW)*, Sep. 2020, pp. 302–309. doi: 10.1109/EuroSPW51379.2020.00047.

[28] J. A. Ingalsbe, L. Kunimatsu, T. Baeten, and N. R. Mead, "Threat Modeling: Diving into the Deep End," *IEEE Softw.*, vol. 25, no. 1, pp. 28–34, Jan. 2008, doi: 10.1109/MS.2008.25.

[29] S. Verreydt, L. Sion, K. Yskout, and W. Joosen, "Relationship-based threat modeling," in *Proceedings of the 3rd International Workshop on Engineering and Cybersecurity of Critical Systems*, in EnCyCriS '22. New York, NY, USA: Association for Computing Machinery, Nov. 2022, pp. 41–48. doi: 10.1145/3524489.3527303.

[30] R. Galvez and S. Gurses, "The Odyssey: Modeling Privacy Threats in a Brave New World," in *2018 IEEE European Symposium on Security and Privacy Workshops (EuroS&PW)*, Apr. 2018, pp. 87–94. doi: 10.1109/EuroSPW.2018.00018.

[31] F. Camelia and T. L. J. Ferris, "Systems Thinking in Systems Engineering," *INCOSE Int. Symp.*, vol. 26, no. 1, pp. 1657–1674, 2016, doi: 10.1002/j.2334-5837.2016.00252.x.

[32] R. Joseph and C. Reigeluth, "The Systemic Change Process in Education: A Conceptual Framework (149)," *Contemp. Educ. Technol.*, vol. 1, pp. 97–117, Apr. 2010, doi: 10.30935/cedtech/5968.

[33] K. Wuyts, D. Van Landuyt, A. Hovsepyan, and W. Joosen, "Effective and efficient privacy threat modeling through domain refinements," in *Proceedings of the 33rd Annual ACM Symposium on Applied Computing*, in SAC '18. New York, NY, USA: Association for Computing Machinery, Apr. 2018, pp. 1175–1178. doi: 10.1145/3167132.3167414.

[34] R. D. Arnold and J. P. Wade, "A Definition of Systems Thinking: A Systems Approach," *Procedia Comput. Sci.*, vol. 44, pp. 669–678, 2015, doi: 10.1016/j.procs.2015.03.050.

[35] M. Frank, "Knowledge, abilities, cognitive characteristics and behavioral competences of engineers with high capacity for engineering systems thinking (CEST)," *Syst. Eng.*, vol. 9, no. 2, pp. 91–103, 2006, doi: 10.1002/sys.20048.

[36] D. Cabrera and L. Cabrera, "What Is Systems Thinking?," in *Learning, Design, and Technology: An International Compendium of Theory, Research, Practice, and Policy*, M. J. Spector, B. B. Lockee, and M. D. Childress, Eds., Cham: Springer International Publishing, 2019, pp. 1–28. doi: 10.1007/978-3-319-17727-4_100-1.

[37] P. M. Senge, *The Fifth Discipline: The Art and Practice of the Learning Organization*. Doubleday/Currency, 2006.

[38] D. Cabrera, L. Cabrera, and E. Powers, "A unifying theory of systems thinking with psychosocial applications," *Syst. Res. Behav. Sci.*, vol. 32, no. 5, pp. 534–545, 2015.

[39] D. Cabrera and L. Cabrera, *Systems thinking made simple: new hope for solving wicked problems*. Ithaca?, New York: Odyssean Press, 2018.

[40] J. C. Davis, P. Amusuo, and J. R. Bushagour, "Experience Paper: A First Offering of Software Engineering," in *Proceedings of The First International Workshop on Designing and Running Project-Based Courses in Software Engineering Education (ICSE-DREE)*, 2022, p. 5.

[41] B. A. Tanay, L. Arinze, S. S. Joshi, K. A. Davis, and J. C. Davis, "An Exploratory Study on Upper-Level Computing Students' Use of Large Language Models as Tools in a Semester-Long Project," in *2024 ASEE Annual Conference & Exposition*, Portland, USA, (Accepted) 2024, pp. 1–27.

[42] A. Greene, "Developing rubrics for open-ended assignments, performance assessments, and portfolios," presented at the Proceedings of the Best Assessment Processes in Engineering Education Conference, Rose-Hulman Institute of Technology, 1997.

[43] D. D. Stevens and A. J. Levi, *Introduction to rubrics: An assessment tool to save grading time, convey effective feedback, and promote student learning*. Routledge, 2023.

[44] S. M. Brookhart and F. Chen, "The quality and effectiveness of descriptive rubrics," *Educ. Rev.*, vol. 67, no. 3, pp. 343–368, Jul. 2015, doi: 10.1080/00131911.2014.929565.

[45] F. McMartin, A. McKenna, and K. Youssefi, "Scenario assignments as assessment tools for undergraduate engineering education," *IEEE Trans. Educ.*, vol. 43, no. 2, pp. 111–119, May 2000, doi: 10.1109/13.848061.

[46] National Institute of Standards and Technology, "The NIST Cybersecurity Framework 2.0," National Institute of Standards and Technology, Gaithersburg, MD, NIST CSWP 29 ipd, 2023. doi: 10.6028/NIST.CSWP.29.ipd.

[47] M. Souppaya, K. Scarfone, and D. Dodson, "Secure Software Development Framework (SSDF) Version 1.1: Recommendations for Mitigating the Risk of Software

Vulnerabilities," National Institute of Standards and Technology, NIST Special Publication (SP) 800-218, Feb. 2022. doi: 10.6028/NIST.SP.800-218.

[48] S. Allen and J. Knight, "A Method for Collaboratively Developing and Validating a Rubric," *Int. J. Scholarsh. Teach. Learn.*, vol. 3, no. 2, Jul. 2009, Accessed: Jan. 31, 2024. [Online]. Available: https://eric.ed.gov/?id=EJ1136714

# Appendix

## Table A1

*STRIDE rubric we developed to assess security case deliverable. Note: We gave a score of 0 if there was no response related to a given construct*

| Constructs | Beginner (score = 1) | Intermediate (Additional to beginner skillset; Score = 2) | Advanced (Additional to intermediate skillset; Score = 3) |
|---|---|---|---|
| *Defining threats in a model* | Identifies if a system has a potential source of threat or not. | Understands the threat that contributes to the risk, and the extent of how the threat impacts the components. | Identify the specific threat agents that can harm the components and/or the system. |
| *Define security requirements* | Identifies critical assets we need to protect (like Confidential Data) | Identify if there are any user roles with varying privileges. | Identify where data can be entered or extracted |
| *Dataflow diagram (DFD)* | 1. Defines trusted boundaries.<br>2. Discusses how the data flows from a no-trusted boundary through other parts of the system | 1. Correctly define trusted boundaries and explain why they are problematic.<br>2. Discusses how the data flows from a no-trusted boundary through other parts of the system.<br>3. Presents a simplistic version of the DFD | 1. Correctly define trusted boundaries and explain why and how they are problematic.<br>2. Describes how the data flows from a no-trusted boundary through other parts of the system.<br>3. Identifies multiple exhaustive problematic regions in a diagram.<br>4. Identifies the extent to which things can go wrong using a more detailed version of DFD |
| *Documentation* | Documents all possible implementations. | Documents the specific strategies to mitigate certain problems. | 1. Discusses how implementation of certain strategies helps with reducing risk to a greater extent than others.<br>2. Discusses ways to reproduce or implement proposed results in detail |
| *STRIDE - Spoofing* | Defines Spoofing and identifies which part of the model contributes to the same | Identifies the property(ies) violated: E.g., Authentication | Identifies the extent to which the threat is affecting the component and/or system and ranks its importance. |
| *STRIDE - Tampering* | Defines Tampering and identifies which part of the model contributes to the same | Identifies the property(ies) violated: E.g., Integrity, Permissions | Identifies the extent to which the threat is affecting the component and/or system and ranks its importance. |
| *STRIDE - Repudiation* | Defines Repudiation and identifies which part of the model contributes to the same | Identifies the property(ies) violated: E.g., Logging, signatures | Identifies the extent to which the threat is affecting the component and/or system and ranks its importance. |
| *STRIDE – Information Disclosure* | Defines Information Disclosure and identifies which part of the model contributes to the same | Identifies the property(ies) violated: E.g., Permission, encryption | Identifies the extent to which the threat is affecting the component and/or system and ranks its importance. |
| *STRIDE – Denial of Service* | Defines Denial of Service and identifies which part of the model contributes to the same | Identifies the property(ies) violated: E.g., Availability | Identifies the extent to which the threat is affecting the component and/or system and ranks its importance. |
| *STRIDE– Elevation of Privilege* | Defines Elevation of Privilege and identifies which part of the model contributes to the same | Identifies the property(ies) violated: E.g., Authorization, sandboxes | Identifies the extent to which the threat is affecting the component and/or system and ranks its importance. |
| *Mitigation strategies* | Identifies correct mitigation strategies based on the properties violated | Discusses the effectiveness of mitigation strategies and specifically answers the questions: Which threat? What strategy? How to implement it? And, | 4. Identifies the threats that exist due to the interaction between the different components and their extent of contribution to the threat of the entire model. Discuss possible mitigation strategies. |

| | | why this implementation mitigates the earlier mentioned problem? | 5. Discusses about efficient implementation of mitigation strategies and talk about resource-constrained situations.<br>6. Discusses possible scope of trade-offs and if there are new requirements in the system for threat mitigation. |
|---|---|---|---|

**Table A2**

*DSRP rubric we developed to assess security case deliverable. Note: We gave a score of 0 if there was no response related to a given construct*

| DSRP Construct | Beginner (score = 1) | Intermediate (Additional to beginner skillset; Score = 2) | Advanced (Additional to intermediate skillset; Score = 3) |
|---|---|---|---|
| *Distinctions* | Identifies basic distinctions between components and system. | Analyzes distinctions in moderately complex threat scenarios. | Critically evaluates and synthesizes complex distinctions between different threats for system and components, showcasing depth. |
| *Systems* | Recognizes basic systems/components in simple scenarios. | Applies DSRP to analyze and synthesize moderately complex systems. Simplify systems enough to be able to analyze different parts and talk about their impact on the greater system. Discusses system-level risks and mitigations. | Applies advanced systems thinking, demonstrating a deep understanding of both component-level and system-level threats using scenarios/discussion/mitigation approaches. |
| *Relationships* | Identifies basic relationships between components. | Analyzes relationships between components in a nuanced manner. | Evaluates intricate relationships between components as well as how they collectively contribute to the system, demonstrating advanced insights. |

| *Perspectives* | Understand basic perspectives within a context (just mentioning insider or outside attackers without discussion or scenarios) | Applies perspectives effectively to analyze threat situations from different attacker standpoints. Does not discuss scenarios of how the attacker might influence the system. Does not distinguish which type of attacker is responsible for which type of risk. | Integrates multiple perspectives of threat situations seamlessly, demonstrating expertise by discussing scenarios of how threats from different perspectives impact the system/components. |
| --- | --- | --- | --- |